\documentclass[reprint,10pt,twocolumn,superscriptaddress,nofootinbib]{revtex4-1}
\usepackage[utf8]{inputenc}
\usepackage{graphicx}
\usepackage{bm}

\usepackage{amsmath,amssymb,amsbsy}
\usepackage{bbm,mathrsfs,yfonts}
\usepackage{float}
\usepackage{hyperref}
\hypersetup{
    colorlinks=true,
    linkcolor=blue,
    filecolor=magenta,      
    urlcolor=cyan,
}
\usepackage{silence}
\usepackage{mathtools}

\renewcommand{\vec}[1]{{\boldsymbol{#1}}}

\newcommand{\gyavg}[1]{\ensuremath{\langle #1 \rangle}}
\newcommand{\gyavgdag}[1]{\ensuremath{\langle #1 \rangle^\dagger}}
\newcommand{\gymom}[1]{\ensuremath{ \big\| #1 \big\|}}
\newcommand{\gymombig}[1]{\ensuremath{ \bigg\| #1 \bigg\|}}

\newcommand{\tnorm}[1]{\ensuremath{ \left| #1 \right|}}

\newcommand{\Bsp}{\ensuremath{B^*_\parallel}} 
\newcommand{\Bs}{\ensuremath{\vec{B}^*}}

\newcommand{\U}[3]{\ensuremath{\prescript{ \scriptscriptstyle #2}{\scriptscriptstyle #3}{\Upsilon}_{#1}}}

\newcommand{\G}[3]{\ensuremath{\prescript{ \scriptscriptstyle #2}{\scriptscriptstyle #3}{\Gamma}_{#1}}}
\newcommand{\Gdag}[3]{\ensuremath{\prescript{\scriptscriptstyle #2}{\scriptscriptstyle #3}{\Gamma}_{#1}^\dagger}}

\newcommand{\Gsq}[3]{\ensuremath{\prescript{ \scriptscriptstyle #2}{\scriptscriptstyle #3}{\check{\Gamma}}_{#1}}}
\newcommand{\Gsqdag}[3]{\ensuremath{\prescript{\scriptscriptstyle #2}{\scriptscriptstyle #3}{\check{\Gamma}}_{#1}^\dagger}}

\newcommand{\Ngy}{\ensuremath{N}}
\newcommand{\Ugy}{\ensuremath{U_\parallel}}
\newcommand{\Ppgy}{\ensuremath{P_\perp}}
\newcommand{\Ppagy}{\ensuremath{P_\parallel}}

\newcommand{\Qp}{\ensuremath{Q_\perp}} 
\newcommand{\Qpa}{\ensuremath{Q_\parallel}} 
\newcommand{\Rpp}{\ensuremath{R_{\perp}}}
\newcommand{\Rpap}{\ensuremath{R_{\parallel,\perp}}} 
\newcommand{\Rpapa}{\ensuremath{R_{\parallel}}} 
\newcommand{\Spp}{\ensuremath{S_{\perp}}}
\newcommand{\Spap}{\ensuremath{S_{\parallel,\perp}}} 
\newcommand{\Spapa}{\ensuremath{S_{\parallel}}}

\newcommand{\ahat}{\hat{\vec a}}
\newcommand{\bhat}{\hat{\vec b}}
\newcommand{\chat}{\hat{\vec c}}

\newcommand{\Fbar}{\ensuremath{{F}}}
\newcommand{\rhogy}{\ensuremath{\varrho}}
\newcommand{\vecrhogy}{\ensuremath{\vec{\varrho}}}
\newcommand{\Zvecbar}{\ensuremath{{\vec{Z}}}}
\newcommand{\Xvecbar}{\ensuremath{{\vec{X}}}}
\newcommand{\Pvecbar}{\ensuremath{{\vec{p}}}}
 
\newcommand{\Zbar}{\ensuremath{{Z}}}
\newcommand{\Xbar}{\ensuremath{{X}}}

\newcommand{\mubar}{\ensuremath{{\mu}}} 
\newcommand{\cpbar}{\ensuremath{{c}_\perp}}
\newcommand{\vpabar}{\ensuremath{{v}_\parallel}}
\newcommand{\cpabar}{\ensuremath{{c}_\parallel}}
\newcommand{\vpatildebar}{\ensuremath{\tilde{{v}}_\parallel}}
\newcommand{\vpabardot}{\ensuremath{\dot{{v}}_\parallel}}
\newcommand{\thetabar}{\ensuremath{{\theta}}} 

\newcommand{\volZbar}{\ensuremath{\text{vol}_{\Zvecbar}}}
\newcommand{\volXbar}{\ensuremath{\text{vol}_{\Xvecbar}}} 
\newcommand{\volPbar}{\ensuremath{\text{vol}_{\Pvecbar}}} 
 
\newcommand{\nablabar}{\ensuremath{\vec{{\nabla}}}} 
\newcommand{\Deltabar}{\ensuremath{{\Delta}}} 
\newcommand{\Gb}[3]{\ensuremath{\prescript{ \scriptscriptstyle #2}{\scriptscriptstyle #3}{{\Gamma}}_{#1}}}

\newcommand{\Gsqb}[3]{\ensuremath{\prescript{ \scriptscriptstyle #2}{\scriptscriptstyle #3}{\check{{\Gamma}}}_{#1}}}

\newcommand{\FullF}{\ensuremath{\text{Full-F }}}
\newcommand{\fullF}{\ensuremath{\text{full-F }}}
\newcommand{\dF}{\ensuremath{\delta\text{F}}}

\newcommand{\Fstat}{\ensuremath{{F_0}}} 
\newcommand{\Ngystat}{\ensuremath{N_0}}
\newcommand{\Tpgystat}{\ensuremath{T_{\perp0}}}
\newcommand{\Ppgystat}{\ensuremath{P_{\perp0}}}
\newcommand{\rhothstat}{\ensuremath{\rho_0}}
\newcommand{\Gstat}[3]{\ensuremath{{\prescript{ \scriptscriptstyle #2}{\scriptscriptstyle #3}{\Gamma}_{#1}}}}
\newcommand{\Gsqstat}[3]{\ensuremath{{\prescript{ \scriptscriptstyle #2}{\scriptscriptstyle #3}{\check{\Gamma}}_{#1}}}}

\newcommand{\gy}{gyro-center }

\newcommand{\gf}{gyro-fluid }
\newcommand{\Gf}{Gyro-fluid }
\newcommand{\AW}{arbitrary perpendicular wavelength }
\newcommand{\SW}{short perpendicular wavelength }
\newcommand{\LW}{long perpendicular wavelength }
\newcommand{\AWs}{arbitrary perpendicular wavelengths }

\newcommand{\AWsdot}{arbitrary perpendicular wavelengths. }
\newcommand{\LWsdot}{long perpendicular wavelengths. }

\newcommand{\ExB}{\ensuremath{\vec{E}\times\vec{B}} }

\newcommand{\curv}{\ensuremath{\vec{\kappa}} }
\newcommand{\opcurv}{\ensuremath{\vec{\mathcal{K}}_{\vec{\kappa}}} }
\newcommand{\opgradb}{\ensuremath{\vec{\mathcal{K}}_{\nablabar B}} }

\begin{document}
\title{
Pad{\'e}-based arbitrary wavelength polarization closures for full-F gyro-kinetic and -fluid models}
\author{M.\ Held}
\email[E-mail: ]{markus.held@chalmers.se}
% \affiliation{Institute for Ion Physics and Applied Physics, 
%                      Universit\"at Innsbruck, A-6020 Innsbruck, Austria}
\affiliation{Department of Space, Earth and Environment, Chalmers University of Technology, SE-412 96 Gothenburg, Sweden}
\author{M.\ Wiesenberger}
\affiliation{Department of Physics, Technical University of Denmark, DK-2800 Kgs. Lyngby, Denmark}
\author{A.\ Kendl}
\affiliation{Institute for Ion Physics and Applied Physics, 
                     Universit\"at Innsbruck, A-6020 Innsbruck, Austria}

\date{\today}
\begin{abstract} %<600 characters including spaces
We propose a solution to the long-standing short wavelength polarization closure shortfall of \fullF  gyro-fluid models. This is achieved by first finding an appropriate quadratic form of the gyro-fluid moment over the polarization part of the \gy  Hamiltonian. Secondly, we deduce Pad{\'e}-based approximations to the latter expression that produce a polarization charge density with the desired order of accuracy and retain linear polarization effects for arbitrary wavelengths. The proposed closures feature proper energy conservation and the anticipated Oberbeck-Boussinesq and long wavelength limits. 
\end{abstract}
\maketitle
\section{Introduction:}
Gyro-fluid  models are an extremely useful approach to provide insights into the behavior of magnetized plasmas.
They are widely applied to study turbulent transport in the tokamak core~\cite{waltz97,staebler05}, edge~\cite{scott00,held19} and scrape-off layer~\cite{madsen11,wiesenberger14,kendl15,held16b,wiesenberger17} and phenomena like collisionless reconnection~\cite{comisso13},  zonal flows~\cite{hahm99,held18} and edge localized modes~\cite{kendl10,xu13}.
\Gf models origin from gyro-kinetic theory~\cite{frieman82,dubin83,lee83} and rely on precise fluid closures to incorporate kinetic effects. They excel due to their vastly reduced computational cost in comparison to gyro-kinetic models and algebraic simplicity in comparison to drift-fluid models since the gyro-viscous cancellations emerge automatically.

\Gf models are particularly characterized by closures, which include finite Larmor radius (FLR) and linear polarization density effects down to the gyro-radius scale~\cite{knorr88,brizard92,hammett92,dorland93}. The latter represent the major hallmarks of gyro-kinetic theory. 
Additionally, \gf closures can also encompass kinetic collisionless dissipation channels like Landau damping or FLR phase mixing~\cite{hammett90,hammett92,dorland93,hunana18}. 

\FullF \gf models~\cite{strintzi04,strintzi05,madsen13}, as opposed to their \(\dF\) counterpart~\cite{knorr88,brizard92,hammett92,dorland93,waltz94,beer96,scott00,snyder01,scott10a}, 
avoid the separation of scales and the concomitant Oberbeck-Boussinesq approximation~\cite{oberbeck1879,boussinesq03}. 
The resulting highly non-linear nature of \fullF  \gf models complicates the development of fluid closures for kinetic effects.
For this reason, polarization in \fullF  \gf models is treated within a long perpendicular wavelength approximation~\cite{strintzi04,strintzi05,madsen13}, which dates back to the beginning of gyro-kinetic theory and neglects polarization at short perpendicular wavelengths~\cite{dubin83}.
This \SW polarization shortfall of current \fullF  \gf models is first of all a fundamental theoretical issue since the \(\dF\) polarization charge density is not recovered in the Oberbeck-Boussinesq limit but only in the additional \LW limit. 
Secondly, the accomplishment of this shortfall is crucial to accurately predict the stability and transport of single- or multi-scale drift wave modes for arbitrary perpendicular wavelengths. 
In particular, \SW structures, which emerge e.g. from ion temperature gradient, trapped electron or interchange modes, can significantly affect the turbulent transport~\cite{hammett92,dorland93,dominski17,mishchenko19,wiesenberger14}. Moreover, non-Oberbeck-Boussinesq polarization effects strongly alter the dynamics of zonal flows and filaments~\cite{held18,kendl15,wiesenberger17}.

%Results
In this contribution, we overcome the \SW polarization shortfall and present novel fluid closures for polarization effects for \AWsdot We find that the exact polarization closure, which rests upon a near Maxwellian distribution function, is not suitable for numerical computations.
Consequently, we refine the latter closure to a Pad{\'e}-based approximation of desired accuracy, which accurately mimics \AW polarization effects in the Oberbeck-Boussinesq limit and retains the original non-linear structure. We prove that the latter is pivotal for energetic consistency - a missing feature of previous ad-hoc approximations~\cite{idomura03,bottino04,ku09,dominski17,mishchenko19}.

%Outline
The remainder of the manuscript is organized as follows. We state the \fullF  gyro-kinetic framework in Sec.~\ref{sec:gk} from which we derive the \fullF  \gf moment hierarchy and Poisson equation in Sec.~\ref{sec:gf}. The \gf closure of the \fullF \gf moment hierarchy and Poisson equation is discussed in Sec.~\ref{sec:gfclosure}, which contains the principal results of the manuscript. In particular, it includes explicit closures for gyro-average effects in Sec.~\ref{sec:FLR}, which enter also the novel closures for polarization effects in Sec.~\ref{sec:pol}. The presented closures exhibit the correct Oberbeck-Boussinesq and \LW limit and retain energetic consistency, which is discussed in Sec.~\ref{sec:OB} and Sec.~\ref{sec:energy}, respectively. Finally, we conclude our findings in Sec.~\ref{sec:conclusion}.
%%%%%%%%%%%%%%%%%%%%%%%%%%%%%%%%%%%%%%%%%%%%%%%%%%%%%%%%%%%%%%%
\section{Gyro-kinetic fundament}\label{sec:gk}
Our discussion is based on the nonlinear electrostatic collisionless gyro-kinetic Vlasov-Poisson system~\cite{frieman82,dubin83,lee83}, which we consistently rederive in this section via field theoretical methods~\cite{sugama00,brizard00,brizard07,scott10a}. We start right after the transformation of the Lagrangian from particle phase-space coordinates \(\vec{z}\) to the \gy phase-space corodinates \(\Zvecbar\), which can be accomplished via Lie-transformation methods. Following the approach developed in Ref.~\cite{sugama00} we define the \gy action
\begin{align}
\label{eq:action}
 S :=& \int dt \sum_\alpha \int  \text{vol}_{\vec{Z}_{0}} \Fbar(\Zvecbar_0,t_0)
 L_{p0}, 
\end{align}
which consists of the \gy particle Lagrangian \(L_p := \gamma_i\dot{\Zbar}^i - H\), a vanishing \gy field Lagrangian and the \gy distribution function \(\Fbar(\Zvecbar,t)\). Here, the \(0\)-subscript denotes  the initial values \((\Zvecbar_0,t_0)\) and we introduced the \gy phase-space coordinates \(\Zvecbar:=\left(\Xvecbar,\Pvecbar\right)\equiv\left(\Xvecbar,\mubar,\vpabar,\thetabar\right)\), which encompass the \gy position \(\Xvecbar\), magnetic moment \(\mubar\), parallel velocity \(\vpabar\) and gyro-angle \(\thetabar\). 
For convenience we omit the species subscript \(\alpha\). 
In the \gy particle Lagrangian the fundamental \gy Poincar\'{e} 1-form and the Hamiltonian are given by 
\begin{align}
\label{eq:poincare}
 \gamma &:= q \vec{A}^* \cdot d\Xvecbar + \frac{m}{q} \mubar d \thetabar, \\
 \label{eq:hamiltonian}
 H&:=\frac{1}{2} m \vpabar ^2 + \mubar  B + q \Psi,
\end{align}
where we introduced the particle charge \(q\) and mass \(m\) and the magnitude \(B:=|\vec{B}|\) of the magnetic field \(\vec{B}\).
In the field part the \gy Hamiltonian of Eq.~\eqref{eq:hamiltonian} appears the central quantity of our work -  the \gy potential \(\Psi :=  \Psi_1 + \Psi_2\). It is the superposition of the gyro-average and polarization contribution of the electric potential \(\phi\),
\begin{align}
\label{eq:Psi12}
 \Psi_1&:= \gyavg{\phi}, &
 \Psi_2&:= - \frac{q}{2 B} \frac{\partial}{\partial \mubar} \left( \gyavg{\phi^2}- \gyavg{ \phi}^2 \right),
\end{align}
respectively. For the definition of the gyro-average \(\gyavg{f} (\Xvecbar,\vpabar,\mubar,t) 
 := \frac{1}{2 \pi } \int_0^{2 \pi} d\thetabar f(\vec{x},\vpabar,t)\) and all therefrom deduced operators a Taylor series representation in configuration space is adopted, since in Fourier space convolution integrals appear, which complicate a simple and clear presentation. 
 Thus, the Taylor series expansion of the gyro-average reads
\begin{align}\label{eq:gyroavg}
 \gyavg{f} (\Xvecbar,\vpabar,\mubar,t) 
   &=\sum_{i=0}^{\infty} \frac{\left(\rhogy^2 \Deltabar_\perp\right)^i}{4^i i!^2} f,
\end{align}
where the coefficients belong to the zeroth Bessel function \(J_0(i x)\) around \(x=0\)~\cite{frieman82,lee83,dubin83}.
For the sake of clarity, we define the particle position \(\vec{x} = \Xvecbar+\vec{\rho}_0 (\Xvecbar,\mubar,\thetabar )\), gyro-radius  \(\rhogy := \sqrt{2 B \mubar / m}/\Omega \), gyro-frequency \(\Omega:=q B / m\), gyro-arm \(\vecrhogy := \rhogy \ahat\) 
and magnetic field unit vector \(\bhat:=\vec{B}/B\). The latter forms with \(\ahat\) and \(\chat\) an orthonormal coordinate system \((\ahat,\bhat,\chat)\). Further, we introduce the abbreviation for the parallel component 
of a vector \(\vec{h}\) with \(h_\parallel = \vec{b}\cdot\vec{h}\) and its perpendicular projection with \(\vec{h}_\perp := -\bhat \times (\bhat \times \vec{h})\).  This defines the perpendicular Laplacian to \(\Delta_\perp := \vec{\nabla}\cdot \vec{\nabla}_\perp\).
\\
In the \gy Poincar\'{e} 1-form of Eq.~\eqref{eq:poincare} the \gy vector potential \(\vec{A}^* := \vec{A} + \frac{m \vpabar}{q} \bhat\) appears, which determines the associated \gy magnetic field \(\Bs := \vec{B} + \frac{m \vpabar}{q} \nablabar \times \bhat\). The parallel component of \(\Bs\) enters the \gy phase-space volume 
\begin{align}
 \volZbar := \sqrt{\text{det}{\left( \omega\right)}} d^6 \Zbar = m^2 |\Bsp| \sqrt{g} d^3 \Xbar d \vpabar \hspace{0.5mm} d \mubar \hspace{0.5mm} d \thetabar,
\end{align}
where \(\omega:=d \gamma\) is the symplectic 2-form and \(g\) is the determinant of the metric tensor. In order to simplify the notation we also define the spatial  and the conjugate momentum space volume 
\begin{align}
\label{eq:volXV}
 \text{vol}_{\Xvecbar}&:= \sqrt{g} d^3 \Xbar, &
 \text{vol}_{\Pvecbar}&:=m^2 |\Bsp|  d \vpabar \hspace{0.5mm} d \mubar \hspace{0.5mm} d \thetabar.
\end{align}
In this work we assume \(\Bs \approx \vec{B} + \frac{m \vpabar}{q}(\nablabar \times \bhat)_\perp\) so that \(\Bsp \approx B \). This simplifies computations and avoids the coordinate singularity at \(\vpabar = -q B/\big[m  (\nablabar\times \bhat )_\parallel\big]\), where without the approximation \(\Bsp =\text{vol}_{\Zvecbar}=0\). 

We now specified all terms induced by the \gy action, which enables us to compute the explicit expression of the \gy Vlasov-Poisson system by means of field theory.
%%%%%%%%%%%%%%%%%%Equations of motion
The \gy equations of motion, which enter the \gy Vlasov equation, can be derived by the variation of the \gy action of Eq.~\eqref{eq:action} with respect to \(\Zvecbar\)~\cite{sugama00}. Constraining this action to be vanishing yields the Euler-Lagrange equation 
\(  \frac{\partial L_p}{\partial \Zbar_i} -\frac{d}{d t} \frac{\partial L_p}{\partial \dot{\Zbar}_i} =0\). The latter can be manipulated to Hamilton's equation \(\frac{d \Zbar^i}{d t} = J^{i j} \frac{\partial H}{\partial \Zbar^j}\), where the components \(J^{ij}\) of the Poisson matrix  \(J:= \omega^{-1}\) emerge. The evaluation of the components of Hamilton's equation 
provides the \gy equations of motion
\begin{subequations}
\label{eq:EOMs}
\begin{align}\label{eq:Xdot}
\dot{\Xvecbar} &= \frac{1}{B} \left(\Bs \vpabar + \frac{\mubar}{q} \bhat \times \nablabar B  + \bhat \times \nablabar \Psi\right),\\
 \label{eq:Vdot}
 \vpabardot &= -\frac{1}{m B} \Bs \cdot \nablabar \left( \mubar B + q \Psi\right), \\
  \label{eq:mudot}
\dot{\mubar}&=0, \\ 
 \label{eq:thetadot}
\dot{\thetabar}&=\Omega + \frac{q \Omega}{B} \frac{\partial}{\partial \mubar} \Psi.
\end{align} 
\end{subequations}
%%%%%%%%%%%%%%%%%%Vlasov equation
The \gy Vlasov equation expresses the conservation of the \gy particle distribution function along the particle trajectories \(\frac{d}{d t}\Fbar=0\). We rearrange this to the conservative form of the \gy Vlasov equation
\begin{align}\label{eq:vlasov}
 \frac{\partial}{\partial t} \left( B \Fbar \right) +\nablabar\cdot \left( \dot{\Xvecbar} B \Fbar \right) + \frac{\partial}{\partial \vpabar} \left( \vpabardot B \Fbar \right)&= 0,
\end{align}
with the \gy equations of motion given by Eqs.~\eqref{eq:EOMs}.
\\
%%%%%%%%%%%%%Poisson part
In gyro-kinetic field theory the \gy Poisson equation follows from the vanishing variation of the \gy action of Eq.~\eqref{eq:action} with respect to \(\phi(\vec{x},t)\), which produces
\begin{align}
\label{eq:Poissonvargy}
 \sum_\alpha \frac{\delta }{\delta \phi(\vec{x},t)} \int \volZbar  \Fbar q \Psi &=0,
\end{align}
Using Eqs.~\eqref{eq:Psi12}  the variation along the electric potential of Eq.~\eqref{eq:Poissonvargy} is explicitly evaluated to
\begin{align}
\sum_\alpha \int\frac{\volPbar}{B}
\left[q \gyavgdag{B \Fbar}+q^2 \left(\phi \gyavgdag{\partial_{\mubar} \Fbar} - \gyavgdag{\gyavg{\phi} \partial_{\mubar} \Fbar}\right)\right] &=0.
\nonumber
\end{align}
Here, we defined 
the adjoint of the gyro-average \(\gyavgdag{ f } (\vec{x},\vpabar,t) := \int \volXbar\gyavg{  f(\Xvecbar,\mubar,\vpabar,t)
 \delta\left(\vec{x} -\Xvecbar-\vec{\rho}_0
 \right)
 } \), since \(\int \volXbar \gyavg{f} h(\Xvecbar) = \int \text{vol}_{\vec{x}} \gyavgdag{h} f(\vec{x})\).
 Note that analogously to the gyro-average of Eq.~\eqref{eq:gyroavg} we can expand the adjoint gyro-average in a Taylor series
 \begin{align}\label{eq:secgyroavg}
\gyavgdag{ f } (\vec{x},\vpabar,t) 
&=  \sum_{i=0}^{\infty}  \frac{\Delta_\perp^i}{4^i i!^2} \left[\rhogy(\vec{x},\mubar)^{2 i} f(\vec{x},\vpabar,t) \right],
\end{align}
 which reveals that the spatial operators act in this case also on the gyro-radius.
%%%%%%%%%%%%%%%%%%%%%%%%%%%%%%%%%%%%%%%%%%%%%%%%%%%%%%%%%%%%%%%
\section{Gyro-fluid moment hierarchy and Poisson equation}\label{sec:gf}
\Gf models are derived by the \gy momentum space integrals over the 
\gy distribution function \(\Fbar(\Zvecbar,t)\) times an arbitrary \gy phase-space function \(\zeta(\Zvecbar,t)\), which defines the \gf moment~\cite{brizard92}
\begin{align}\label{eq:gyromom}
 \gymom{\zeta} &:= \int \volPbar \Fbar \zeta.
\end{align}
The latter \gf moment is exploited for the basic \gf moment quantities
\begin{subequations}
\label{eq:momentvars}
\begin{align}
  \Ngy &:=\gymom{1},                  &   
  \Ugy&:=\gymom{\vpabar}/\Ngy, \\
  \Ppgy&:= \gymom{\mubar B }, &
  \Ppagy&:=\gymom{m \vpatildebar^2  },  \\
  \Qp &:=\gymom{\mubar B  \vpatildebar }, &
 \Qpa &:=\gymom{m\vpatildebar^3},
\end{align}
\end{subequations}
the  \gy density, parallel velocity and the perpendicular and parallel pressure and the perpendicular and parallel contributions of the parallel heat flux, respectively.
Here, we defined \(\vpatildebar := \vpabar - \Ugy\). Additionally we relate the perpendicular and parallel pressure to the corresponding temperature via the ideal gas law  \(\Ppgy = \Ngy T_\perp \) and \(\Ppagy= \Ngy T_\parallel\).
%%%%%%%%%%%%%Moment hierarchy%%%%%%%%%%%%%%%%%%%%%%%%%%%%%%%%
Further, the \gf moment over the Vlasov-equation~\eqref{eq:vlasov} multiplied by \(\zeta\) yields the general expression for the time 
evolution of the \gf moments~\cite{brizard92}
\begin{align}\label{eq:gfmomevo}
\frac{\partial }{\partial t} \gymom{\zeta }+
\nablabar\cdot  \gymom{\zeta \dot{\Xvecbar} } 
&= \gymombig{\frac{d }{d t}\zeta}
\end{align}
with \(\frac{d}{d t} \zeta = \left(\frac{\partial }{\partial t} +\dot{\Xvecbar} \cdot \nablabar+  \vpabardot \frac{\partial}{\partial \vpabar}\right)\zeta\).
%%%%%%%%%%%%Explicit moment equations
Inserting the \gy equations of motions of Eqs.~\eqref{eq:EOMs} into the \gf moment evolution Eqs.~\eqref{eq:gfmomevo} 
and restricting the phase space function to \(\zeta = \mubar^k \vpabar^l\) yields the \gf moment hierarchy evolution equation
\begin{align}
\label{eq:gfmomevomuv}
\frac{\partial }{\partial t} &\gymom{\mubar^k \vpabar^{l} }
+
\nablabar\cdot \bigg(
 \bhat\gymom{\mubar^k \vpabar^{l+1} } 
+\frac{m \opcurv}{q}  \gymom{\mubar^k \vpabar^{l+2} }  
\nonumber \\
&
+\frac{B \opgradb}{q} \gymom{\mubar^{k+1} \vpabar^l }
+\frac{\bhat }{B} \times\gymom{\mubar^k \vpabar^l  \nablabar\Psi }  \bigg)
\nonumber \\
=&
-l\bigg(
\frac{q}{m} \bhat \cdot\gymom{\mubar^k \vpabar^{l-1}   \nablabar \Psi }
+ \frac{\bhat \cdot \nablabar B}{m}\gymom{ \mubar^{k+1} \vpabar^{l-1}  }
\nonumber \\
&+  \opcurv  \cdot \gymom{ \mubar^k \vpabar^l \nablabar \Psi }
+ \frac{ \opcurv  \cdot \nablabar B}{q} \gymom{\mubar^{k+1} \vpabar^l 
}   \bigg) .
\end{align}
Here, we defined the curvature \(\curv := \bhat \cdot \nablabar \bhat\) and the \(\nablabar B\) drift related terms \(\opcurv:=\frac{1}{B} \bhat \times \curv\) and \(\opgradb:=\frac{1}{B} \bhat \times \nablabar \ln{(B)}\). 
The moment hierarchy evolution Eqs.~\eqref{eq:gfmomevomuv} encompass in general an infinite set of evolution equations for all non-negative integer powers of \(\gymom{\mubar^k \vpabar^{l} }\). We target the formulation of a six moment \gf model for the basic \gf moment quantities of Eq.~\eqref{eq:momentvars} so that \(k\leq1\) and \(l\leq 3\). As a consequence we need to close in Eq.~\eqref{eq:gfmomevomuv} not only the higher \gf moment quantities
\begin{subequations}
\label{eq:highermomentvars}
\begin{align}
 \Rpp        &:= \gymom{(\mubar B)^2 }, &
 \Rpap    &:= \gymom{\mubar B m \vpatildebar^2  }, \\
 \Rpapa&:=\gymom{m^2 \vpatildebar^4  }, &
 \Spap &:=\gymom{ \mubar B  m \vpatildebar^3  },\\
 \Spp     &:=\gymom{(\mubar B)^2  \vpatildebar },&
 \Spapa   &:=\gymom{ m^2\vpatildebar^5  },
\end{align}
\end{subequations}
but also \(\Psi\) related terms of the form \( \gymom{ \mubar^k \vpabar^l \nablabar\Psi }\). 

%%%%%%%%%%%%Poisson equation %%%%%%%%%%%%%%%%%%%%%%%%%%%%%%%%%%
The remaining part of the \gf model is the \gf Poisson equation.
A crucial move in the deduction of \gf closures for gyro-average and polarization effects is to express the \gy Poisson Eq.~\eqref{eq:Poissonvargy} in terms of the \gf moment \(\gymom{\Psi}\) before the variation along \(\phi\) is explictly evaluated. 
Accordingly, we express the \gy Poisson equation in terms of the \gf moment \(\gymom{\Psi}\) and split 
it into a \gy charge density \( \sum_\alpha  q \Ngy\) and a polarization charge density \( -\vec{\nabla} \cdot \vec{P}\) contribution
 \begin{align}
 \label{eq:gfpoisson}
 \sum_\alpha  q \Ngy -  \vec{\nabla}\cdot\vec{P}&=0.
\end{align}
The polarization density \(\vec{P}\) can be decomposed into first and second order polarization contributions \(\vec{P} = \vec{P}_1 +  \vec{P}_2 \), which are associated with \(\Psi_1\) and \(\Psi_2\), respectively.
The first order polarization density
\begin{align}
\label{eq:NFLRdef}
\vec{P}_1 := \sum_\alpha \vec{\nabla}_\perp \int \volXbar q \frac{\delta  \gymom{\phi-\Psi_1}}{\delta (\Delta_\perp \phi(\vec{x},t))} 
\end{align}
contains FLR effects and is independent of the electric potential \(\phi\). 
The second order polarization density 
\begin{align}
   \label{eq:Npoldef2}
 \vec{P}_2 &:=\sum_\alpha \int \volXbar q \frac{\delta \gymom{\Psi_2 } }{\delta (\vec{\nabla}_\perp \phi(\vec{x},t))} , 
\end{align}
can be rewritten into a linear polarization density \(\vec{P}_2 = \epsilon_0 \mathbb{X} \cdot \vec{\nabla}_\perp \phi \) and consequently contributes only for a non-vanishing electric field~\cite{krommes93}. 
Here, \(\epsilon_0\) is the vacuum permittivity and  \(\mathbb{X}\) is the rank-2 electric susceptibility tensor, which contains 2nd order polarization effects for \AWsdot
From now on we associate first and second order polarization contributions with gyro-average (or FLR) and polarization effects, respectively.

The \gf moment hierarchy evolution Eqs.~\eqref{eq:gfmomevomuv} and the Poisson equation~\eqref{eq:gfpoisson} demand a closure for terms of the form \( \gymom{ \mubar^k \vpabar^l \nablabar\Psi }\) and \(\gymom{\Psi}\). 
The accurate evaluation of the closure terms \( \gymom{ \mubar^k \vpabar^l \nablabar\Psi }\) and \(\gymom{\Psi }\) 
needs an infinite set of \gf moment quantities, even if a closed set of evolution equations, as for instance for the basic \gf moment quantities of Eq.~\eqref{eq:momentvars} with \(k\leq1\) and \(l\leq 3\), is calculated.
This peculiarity of \gf models is reasoned in the gyro-average of Eq.~\eqref{eq:gyroavg}, which entails all non-negative integer powers of the \gy magnetic moment \(\mubar\)~\cite{dorland93}. In particular, the \gf moment over the gyro-average and polarization part of the \gy potential \(\Psi\) can be Taylor expanded according to
\begin{subequations}
\label{eq:psiarb}
\begin{align}
 \label{eq:psi1arb}
   \gymom{\Psi_1} =& \left( \Ngy + \frac{\Ppgy\Deltabar_\perp }{ 2 m \Omega^2 }  +  \frac{\Rpp \Deltabar_\perp^2}{16 m^2 \Omega^4 }   + \dots\right)\phi,
   \\
\label{eq:psi2arb}
 \gymom{\Psi_2} 
 =&  - \frac{q \Ngy}{2 m \Omega^2 } \tnorm{\nablabar_\perp\phi }^2 
       - \frac{q \Ppgy }{ 8 m^2 \Omega^4 } \Big[ 2 | \nablabar_\perp^2 \phi|^2
       \nonumber \\&
        + 4 \nablabar_\perp \phi\cdot \Deltabar_\perp \nablabar_\perp \phi -\left(\Deltabar_\perp \phi \right)^2 \Big] 
 \nonumber \\
 &-\frac{q \Rpp}{48 m^2 \Omega^4 }   \Big[2 \tnorm{\nablabar_\perp^3\phi }^2 
        + 3\nablabar_\perp  \phi \cdot \Deltabar_\perp^2 \nablabar_\perp \phi
\nonumber \\
&+  3\tnorm{\Deltabar_\perp\nablabar_\perp  \phi }^2+6 \nablabar_\perp^2\phi : \Deltabar_\perp\nablabar_\perp^2\phi
\nonumber \\
&-3\Deltabar_\perp \phi \Deltabar_\perp^2 \phi\Big]
 +\dots
\end{align}
\end{subequations}
Here, the norm of a rank-n tensor \(\mathbb{T}\) is defined by \(\tnorm{\mathbb T} := \sqrt{\mathbb{T}_{i_1,\dots,i_n}\mathbb{T}^{i_1,\dots,i_n}}\) and \(\nablabar_\perp^i\) represents \((i-1)\)-times a tensor product, e.g. for \(i=2\) we get \(\nablabar_\perp^2 = \nablabar_\perp \nablabar_\perp\).
\section{Gyro-fluid closure}\label{sec:gfclosure}

In \fullF  theory so far two different strategies are pursued to resolve this infinite hierarchy, which is triggered by 
the \( \gymom{ \mubar^k \vpabar^l \nablabar\Psi }\) and \(\gymom{\Psi}\) terms. 
The first most trivial approach truncates the Taylor series expansion at a specific order and opts usually towards a \LW approximation, so that \(\gymom{\Psi_1} \approx \left( \Ngy + \frac{\Ppgy\Deltabar_\perp }{ 2 m \Omega^2 } \right)\phi\) and \(\gymom{\Psi_2} 
 \approx - \frac{q \Ngy}{2 m \Omega^2 } \tnorm{\nablabar_\perp\phi }^2 \). However, such approximated \gf models are not advantageous over drift-fluid models, since they treat the FLR and polarization terms with the same level of detail. As opposed to this, the second approach retains \AW effects by restricting the \gy distribution function \(\Fbar\) to a specific form, e.g. a Maxwellian as was originally used by Refs.~\cite{knorr88,brizard92}. In general, this approach truncates the Laguerre-Hermite expansion of the \gy distribution function \(\Fbar\) at a particular polynomial order.
As a consequence, FLR and polarization effects are kept for \fullF  to a specific order of the Taylor series expansion and for truncated \(\Fbar\) above this order.
\\
In \(\dF\) theory also other closure strategies can be utilized, which rest upon the linear \gy solution of the Vlasov equation~\cite{hammett92,dorland93}. However, such methods are inapplicable in \fullF  theory, since \fullF  closures must avoid both the splitting of the \gy distribution function  \(\Fbar=\Fstat(1 +\delta F)\) into a stationary Maxwellian \(\Fstat\) and small fluctuating non-Maxwellian \(\delta F\) part and the Oberbeck-Boussinesq approximation.

We follow now the second approach and evaluate the \gf moments over the various closure terms \( \gymom{ \mubar^k \vpabar^l \nablabar\Psi }\) and \(\gymom{\Psi }\) to \AWs by specifing the \gy distribution function \(\Fbar\).
First we expand the \gy distribution function in a Laguerre-Hermite polynomial in \((\mubar,\vpatildebar)\) space
~\cite{jorge17,mandell18,frei19}
\begin{align}
\label{eq:FHL}
 \Fbar = \Fbar_M \sum_{i=0}^\infty\sum_{j=0}^\infty \mathcal{N}_{i j}  L_i (\cpbar) H_j (\cpabar)  ,
\end{align}
with Maxwellian 
 \begin{align}
 \label{eq:FM}
 \Fbar_M&:= \frac{\Ngy}{2 \pi  T_\perp m} \sqrt{\frac{1}{2 \pi  T_\parallel m}} \exp{\left( -  \cpbar -\cpabar^2\right)}, 
\end{align}
expansion coefficients \(\mathcal{N}_{ij}\), \(\cpabar :=\vpatildebar\sqrt{m/(2 T_\parallel)}\) and \(\cpbar:= \mubar B/T_\perp\).
Here, we introduced the Laguerre polynomial 
\(L_i(x) := \frac{\exp{\left({x}\right)}}{i!} \frac{d^i}{d x^i} \left[\exp{\left({-x}\right)} x^i\right] \) and the physicist's Hermite polynomial 
\(H_j(x) := (-1)^j \exp{\left({x^2}\right)} \frac{d^j}{d x^j} \exp{\left({-x^2}\right)}\). 
Secondly, we approximate the \gy distribution function by truncating the Laguerre-Hermite expansion at order \(\left(I,J\right)\) so that 
\begin{align}
\label{eq:FIJ}
 \Fbar \approx \Fbar_M  \sum_{i=0}^I\sum_{j=0}^J \mathcal{N}_{i j}   L_i (\cpbar) H_j(\cpabar) =: \Fbar_{IJ}.
\end{align}
This is known as closure by truncation, since all expansion coefficients above \(\left(I,J\right)\)  are set to zero. However, this approach does not rule out polynomial or asymptotic closures for smaller \(\left(I,J\right)\). 
For the sake of clarity we define also the truncated \gf moment
\begin{align}
\label{eq:gyromomIJ}
 \gymom{\zeta}_{IJ} :=\int \volPbar \Fbar_{IJ} \zeta,
\end{align}
which rests upon the truncated \gy distribution function of Eq.~\eqref{eq:FIJ}.

For the remainder of the manuscript we truncate the \gy distribution function of Eq.~\eqref{eq:FIJ} at \(\left(I,J\right)=\left(1,3\right)\) and fix the only non-vanishing expansion coefficients to \(\mathcal{N}_{00} := 1\), \(\mathcal{N}_{11} := -\frac{ \Qp}{  \Ppgy } \sqrt{\frac{m}{2 T_\parallel}}\) and  \(\mathcal{N}_{03} := \frac{ \Qpa}{12  \Ppagy} \sqrt{\frac{m}{2 T_\parallel}}\).
 This results in a near Maxwellian distribution function \(\Fbar_{13}\), which could be written as in Ref.~\cite{madsen13}
\begin{align}
\label{eq:F13}
 \Fbar_{13} &=\Fbar_M  \left(1+ \xi\right) ,  
\end{align}
 with \(
 \xi :=   \sqrt{\frac{2  m}{ T_\parallel}} \cpabar \left[\frac{\Qp}{ \Ppgy } \left(\cpbar - 1\right)  + \frac{\Qpa}{6 \Ppagy } 
 \left(2\cpabar^2-3\right)\right]
 \).
As a consequence of the near Maxwellian assumption of Eq.~\eqref{eq:F13} the \gf hierarchy is closed by truncation, because the higher \gf moment quantities of Eq.~\eqref{eq:highermomentvars}  are expressed through the basic \gf moments
\begin{subequations}
\label{eq:highermomentvarsclosed}
\begin{align}
 \Rpp     &= 2 \Ppgy T_\perp, &
 \Rpap    &= \Ppagy T_\perp , \\ 
 \Rpapa   &= 3 \Ppagy T_\parallel ,  &
 \Spap    &=3 \Qp T_\parallel + \Qpa T_\perp ,\\ 
 \Spp     &= 4 \Qp T_\perp  ,&
 \Spapa   &= 10 \Qpa T_\parallel.
\end{align}
\end{subequations}
In certain cases a different closure, such as a polynomial or asymptotic (e.g. collisional) closure, is intended instead of the closure by truncation of Eq.~\eqref{eq:highermomentvarsclosed}.
Then one must either resort 
to a four moment model with \((\Ngy,\Ugy,\Ppgy,\Ppagy)\), where the chosen closure is determined by \((\Qp,\Qpa)\), 
or adopt the higher truncation  \(\left(I,J\right)=\left(2,5\right)\), so that \((\Rpp, \Rpap, \Rpapa, \Spap, \Spp, \Spapa)\) retake the role of unspecified closure variables. 
However, the gyro-average and polarization closures are increasingly complex to implement numerically for higher truncations, such as \(\left(I,J\right)=\left(2,5\right)\), than those at \(\left(I,J\right)=\left(1,3\right)\).
Thus, from the practical point of view it is reasonable to close the higher gyro-fluid moment quantities in the gyro-average and polarization terms differently than in the remaining \gf model, even though full consistency within the \gf model is lost. 
The derivation of consistent full-F gyro-fluid closures for e.g. \((\Qp,\Qpa)\) at arbitrary collisionalities is an ongoing effort, which will be reported in a future work.

The near Maxwellian assumption of Eq.~\eqref{eq:F13} allows us to evaluate the various \gf moment closure terms \(\gymom{\mubar^k \vpabar^l \nablabar\Psi}\) and \(\gymom{ \Psi}\), which generate a hierarchy of \gf potentials. 
These \gf moment closure terms are evaluated for a six moment \gf model to 
\begin{subequations}
\label{eq:gfmomclosures}
\begin{align}
\label{eq:nablaPsimom}
 \gymom{ \nablabar \Psi}_{13}  =& \Ngy \left( \nablabar\psi +  \chi  \nablabar \eta \right), \\
%%%%%%%%%%%%
\label{eq:muBnablaPsimom}
 \gymom{\mubar B \nablabar \Psi}_{13}  =&
 \Ppgy \left[\nablabar \left(\psi + \chi\right) + \iota \nablabar\eta \right], 
 \\
 %%%%%%%%%%%%
\label{eq:vnablaPsi2mom}
  \gymom{\vpabar \nablabar \Psi}_{13} =&
  \Ngy \Ugy \nablabar \psi +  \frac{\Qp}{T_\perp} \nablabar \chi 
    \nonumber \\&
    + \chi \left( \Ngy \Ugy - \frac{\Qp}{T_\perp} \right)\nablabar \eta 
  +  \frac{\Qp}{T_\perp} \iota \nablabar \eta,  \\
 %%%%%%%%%%%%
 \label{eq:mv2nablaPsi2mom}
 \gymom{m \vpabar^2 \nablabar \Psi}_{13} =& 
 \left(\Ppagy +m \Ngy \Ugy^2 \right) \nablabar \psi + \frac{2 m \Ugy \Qp}{T_\perp}\nablabar \chi 
 \nonumber \\&
 + \left(\Ppagy + m \Ngy \Ugy^2 - \frac{2 m \Ugy \Qp}{T_\perp} \right) \chi \nablabar \eta 
 \nonumber \\  &
 + \frac{2 m \Ugy \Qp}{T_\perp} \iota \nablabar \eta,  \allowdisplaybreaks 
 \\
 %%%%%%%%%%%%%%%%%%%%
 \label{eq:mv3nablaPsi2mom}
 \gymom{m \vpabar^3 \nablabar \Psi}_{13} =& 
 \left(\Qpa +m \Ngy \Ugy^3 +3 \Ppagy \Ugy\right) \nablabar \psi 
 \nonumber \\  &
 +  \left(\frac{\Spap+3 m \Ugy^2 \Qp}{T_\perp}- \Qpa \right) \nablabar \chi 
 \nonumber \\ &
 + \bigg(2 \Qpa + 3 \Ugy \Ppagy + m \Ngy \Ugy^3
 \nonumber \\  &- \frac{\Spap+3 m \Ugy^2 \Qp}{T_\perp} \bigg) 
  \chi \nablabar \eta  
 \nonumber \\ &
 +  \left(\frac{\Spap+3 m \Ugy^2 \Qp}{T_\perp}- \Qpa \right) \iota \nablabar \eta,  
 \\
 %%%%%%%%%%%%%%%%%%%%
 \label{eq:vmuBnablaPsi2mom}
\gymom{\vpabar \mubar B \nablabar \Psi}_{13} =& 
\left( \Ppgy \Ugy +\Qp\right) \nablabar \left(\psi + \chi\right) + \Qp \nablabar \iota 
\nonumber \\&
+ \left(\Ppgy \Ugy - \Qp \right) \iota \nablabar \eta + \Qp  \upsilon\nablabar \eta,
\end{align}
\end{subequations}
with \(\eta := \ln{(B/T_\perp)}\).
The \gf moment closure terms of Eqs~\eqref{eq:gfmomclosures} produce the basic \( \psi:=\psi_1 + \psi_2 \)
and higher  \gf potentials \( \chi:=\chi_1 + \chi_2 \) and \( \iota:=\iota_1 + \iota_2\). The gyro-fluid potentials \((\psi,\chi,\iota)\) arise from the \gf moment over the \gy potential \(\Psi\). The respective gyro-average and polarization parts of the gyro-fluid potentials are defined by
\begin{subequations}
\label{eq:psichiiotaupsilon12def}
\begin{align}
\label{eq:psi12def}
 \psi_i&:=\gymom{\Psi_i }_{13} /\Ngy, \\
\label{eq:chi12def}
 \chi_i&:=\gymom{(\mubar  B/T_\perp -1) \Psi_i }_{13} /\Ngy, \\
\label{eq:iota12def}
 \iota_i&:=\gymom{\mubar  B/T_\perp(\mubar  B/T_\perp-2) \Psi_i }_{13} /\Ngy,
\end{align}
\end{subequations}
with \(i \in \left\{1,2\right\}\). This hierarchy retains gyro-average (\(\Psi_1\)) as well as polarization (\(\Psi_2\)) effects through \(\left(\psi_1,\chi_1,\iota_1\right)\) and \(\left(\psi_2,\chi_2,\iota_2\right)\), respectively.
\\
The gyro-average of Eq.~\eqref{eq:gyroavg} is inherent to both the closure for gyro-average and polarization effects (cf.~\eqref{eq:Psi12}).
As a consequence, the higher \gf potentials \(\left(\chi,\iota\right)\)  of not only the gyro-average \(\left(\chi_1,\iota_1\right)\) but also of the polarization \(\left(\chi_2,\iota_2\right)\) can be related to the basic \gf potential \(\psi\) by the simple and exact recursive identities~\cite{dorland93}
\begin{subequations}
\label{eq:Dorlandidentities}
\begin{align}
 \chi_i &= \frac{\rho}{2} \frac{\partial }{\partial \rho} \psi_i, \\
 \iota_i &= 
\left(1 +  \frac{\rho}{2} \frac{\partial }{\partial \rho} \right)\chi_i, & i \in \left\{1,2\right\}.
 \end{align}
 \end{subequations}
 Here, we defined the thermal gyro-radius \(\rho := \sqrt{T_\perp m}/(q B)\). These formulas provide consistent closures for the gyro-average and polarization contributions of Eqs.~\eqref{eq:chi12def} and~\eqref{eq:iota12def}, even if an approximated closure of the lowest moment \(\psi\) is utilized. 
This result extends the recursive identities, originally derived for the gyro-average contributions~\cite{dorland93}, to the polarization contributions. 
%%%%%%%%%%%%Gyro-average part %%%%%%%%%%%%%%%%%%%%%%%%%
\subsection{Gyro-average closures}\label{sec:FLR}
The gyro-average contributions in the \gf potentials \(\psi_1\) and \(\chi_1\), \(\iota_1\) give rise to the basic and higher FLR operators, \(\G{1}{}{}\) and \(\G{2}{}{}\), \(\G{3}{}{}\), respectively. 
These FLR operators are defined via the \gf potentials according to
\begin{align}
 \Gb{1}{}{}(\phi) &:=\psi_1, & 
 \Gb{2}{}{}(\phi) &:=\chi_1, &
 \Gb{3}{}{}(\phi) &:=\iota_1 . 
\end{align}
The near Maxwellian distribution function of Eq.~\eqref{eq:F13} allows us to evaluate these operators for \AWsdot 
The consequent basic FLR operator \( \G{1}{}{}\) is given in configuration space by a Taylor series
\begin{align}
\label{eq:gamma1}
 \G{1}{}{} &= \sum_{i=0}^{\infty} \frac{\left(\rho^{2}  \Delta_\perp\right)^i }{2^{i} i!}, 
 \end{align}
with the familiar Taylor series coefficients of the exponential \( \exp{(x^2/2)} \) at \(x=0\)~\cite{knorr88,brizard92}. 
Note that Eq.~\eqref{eq:gamma1} renders the Taylor expansion for \fullF  of Eq.~\eqref{eq:psi1arb} up to \(\mathcal{O}(b^2)\) with \(b:=\rho k_\perp\).
As a result it features \fullF  FLR effects up to \(\mathcal{O}(b^2)\) and near Maxwellian FLR effects for all higher even orders of \(b\).
\\
As soon as the basic FLR operator is determined either by the near Maxwellian assumption (Eq.~\eqref{eq:gamma1}) or by an additional approximation (discussed later), the higher FLR operators and the first order polarization (FLR) contribution in the Poisson equation follow immediately. 
For the higher FLR operators \(\G{2}{}{}\) and \(\G{3}{}{}\) the recursive identities of Eqs.~\eqref{eq:Dorlandidentities} are utilized in combination with Eq.~\eqref{eq:gamma1}. 
This yields \(\G{2}{}{}= \sum_{i=0}^{\infty} \frac{(\rho^{2}  \Delta_\perp)^i}{2^{i} (i-1)!} \) and 
\(\G{3}{}{}=  \sum_{i=0}^{\infty} \frac{(1+i) (\rho^{2}  \Delta_\perp)^i }{2^{i} (i-1)!} \). 
The first order polarization charge density is produced 
by the adjoint \(\Gdag{1}{}{}\) of the basic FLR operator~\cite{strintzi04,madsen13}
\begin{align}
-\vec{\nabla} \cdot \vec{P}_1 &= \sum_\alpha q \left(\Gdag{1}{}{}-1\right) \Ngy.
\end{align} 
and emerges from the variationial of Eq.~\eqref{eq:NFLRdef}.

While the latter FLR operators accurately capture gyro-averaging effects for the near Maxwellian distribution function, they fail to match the linear gyro-kinetic solution and consequently the 
ion temperature gradient marginal stability relation for a finite set of gyro-moments in a slab and constant magnetic field~\cite{hammett92,dorland93,mandell18}. 
Therefore, approximations to Eq.~\eqref{eq:gamma1} are needed, which are well behaved and capture both gyro-averaging effects and the linear response and consequently the ITG instability properly. 

The \(\sqrt{\Gamma_0}\) approximation overcomes this drawback~\cite{hammett92,dorland93} and 
replaces the basic FLR operator by 
\begin{align}
\label{eq:gamma1012}
 \G{1}{}{} &\approx \sqrt{\Gamma_0} =: \Gsq{1}{}{}.
\end{align}
Here, the linear polarization operator is defined by
\begin{align}
 \label{eq:gamma0}
  \G{0}{}{} &:=\sum_{i=0}^{\infty}  \frac{ 2^i \Gamma(i+1/2)}{\sqrt{\pi} \left(i!\right)^2} \left(\rho^{2}\Delta_\perp\right)^i,
\end{align}
where the Gamma function \(\Gamma\) is not to be mistaken with the gyro-average or polarization operator.
It originates from  the Oberbeck-Boussinesq limit of the Poisson equation \(\G{0}{}{}(\phi) = \gymom{\gyavgdag{\gyavg{\phi}}}_{13}/\Ngy\), which we relax after integration for the thermal gyro-radius \(\rho\). 
The Taylor series coefficients of the linear polarization operator \(\G{0}{}{}\) correspond to the well-known function \(I_0(-x^2)\exp{(x^2)}\) at \(x=0\)~\cite{dubin83,lee83}, where \(I_n\) is the modified Bessel function.

\begin{figure}[ht]
\centering
\includegraphics[width= 0.49\textwidth]{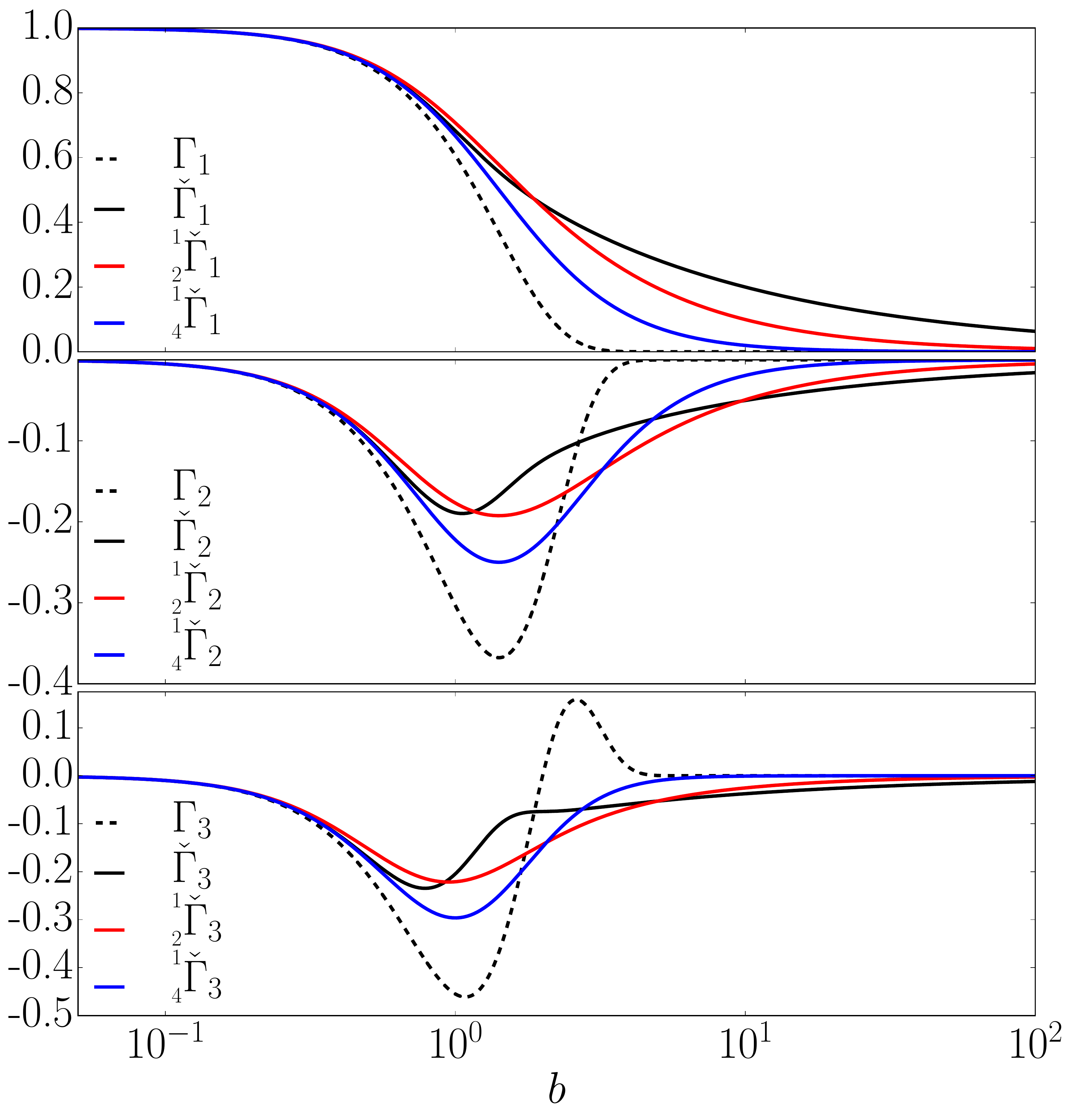}
\caption{The near Maxwellian FLR operators \(\G{1}{}{}\),\(\G{2}{}{}\) and \(\G{3}{}{}\) (dashed), their square root approximation \(\Gsq{1}{}{}\),\(\Gsq{2}{}{}\) and \(\Gsq{3}{}{}\) (black) and the corresponding \((1,2)\) and \((1,4)\) Pad{\'e}-approximations (red, blue) are shown.}
\label{fig:gamma1234approx}
\end{figure}

From the numerical point of view neither the near Maxwellian basic FLR operator (Eq.~\eqref{eq:gamma1}) nor its \(\sqrt{\Gamma_0}\) approximation (Eq.~\eqref{eq:gamma1012}) are practical in \fullF  \gf models. This is because in configuration space accuracy to \AWs is lost due to a truncation of the Taylor series and in Fourier space computationally demanding convolution integrals emerge.
Pad{\'e}-approximations of order \((P,Q)\) in configuration space offer a way out of this difficulty. 
The \((P,Q)\) Pad{\'e}-approximation is a uniquely determined rational approximation to a function or in general an operator \(\U{}{}{}\), which we define as
\begin{align}
\U{}{P}{Q}(x) \approx \frac{ \sum_{i=0}^P \alpha_i x^i}{1+ \sum_{i=0}^Q \beta_i x^i}
\end{align}
Here, the series coefficients \(\alpha_i\) and \(\beta_i\) are determined from the condition \(\U{}{}{}^{(P+Q)}(0) =  \left(\U{}{P}{Q} \right)^{(P+Q)}(0)\). In this work these coefficients are deduced by the \textit{PadeApproximant} routine of \textit{Wolfram Mathematica}~\cite{Mathematica}.
%%%%%%%%
Proper approximations to the basic FLR operator \(\Gamma_1\) are based on a suitable \((P,Q)\) Pad{\'e}-approximation, abbreviated by \(\U{}{P}{Q}\), 
to the chosen operator \(\U{}{}{}\). 
For the gyro-average part of the basic \gf potential we choose for \(\U{}{}{}\) the polarization operator \(\Gamma_0\) (or its square root) whereas for the polarization part of the basic \gf potential, as we discuss later, we pick the basic FLR operator \(\Gamma_1\) (or its square).
Two \(\mathcal{O}(b^2)\) accurate and at \(b\rightarrow \infty\) vanishing Pad{\'e}-approximations  for the basic FLR operator  \(\Gsq{1}{P}{Q}\) emerge at order \((1,2)\)~\cite{hammett92}
\begin{align}
  \G{1}{}{} &\approx  
\sqrt{\G{0}{1}{2}} =\sqrt{\prescript{\scriptscriptstyle 1}{\scriptscriptstyle 2}{\left(\G{1}{}{}^2\right)}} = \frac{1}{\sqrt{1-\rho^2\Delta_\perp}}=: \Gsq{1}{1}{2}, 
\end{align}
and \((1,4)\)~\cite{dorland93,strintzi04,madsen13}
\begin{align}
  \G{1}{}{} &\approx 
 \prescript{\scriptscriptstyle 1}{\scriptscriptstyle 2}{\sqrt{\Gamma}_{0}} = \G{1}{1}{2} = \frac{1}{1-\rho^2/2\Delta_\perp}=: \Gsq{1}{1}{4},
\end{align}
Note that for the latter \((1,2)\) and \((1,4)\) Pad{\'e}-approximation 
the \(\sqrt{\G{0}{}{}}\) relationship of Eq.~\eqref{eq:gamma1012} is exactly fulfilled, so that \(
\prescript{\scriptscriptstyle 1}{\scriptscriptstyle 2}{\left(\G{1}{}{}^2\right)} = \G{0}{1}{2} \) 
and \( \G{1}{1}{2} =  \prescript{\scriptscriptstyle 1}{\scriptscriptstyle 2}{\sqrt{\Gamma}_{0}} \), respectively.
The higher Pad{\'e} approximated FLR operators follow from the recursive identities of Eq.~\eqref{eq:Dorlandidentities} and are summarized together with the basic  Pad{\'e} approximated FLR operator and its adjoint in Table~\ref{table:PadeappstoFLRops}~\cite{held16b}.
\begin{table*}[ht]
 \caption{Pad{\'e}-approximations for FLR operators}
 \begin{ruledtabular}
  \label{table:PadeappstoFLRops}
 \begin{tabular}{ c  c  c   c  c  }   
  \(q\) & \(\Gsq{1}{1}{q}\) &\(\Gsqdag{1}{1}{q}\) & \(\Gsq{2}{1}{q}\) & \(\Gsq{3}{1}{q}\)\\
    \hline
  \(2\) & \(\frac{1}{\sqrt{1-\rho^2 \Delta_\perp}}\) & \(\frac{1}{\sqrt{1- \Delta_\perp\rho^2}}\) & \(\frac{\rho^2}{2} \frac{1}{\sqrt{\sum_{i=0}^{3} \binom{3}{i}(-\rho^{2} \Delta_\perp)^i}}\Delta_\perp\) & \(\rho^2 \frac{1}{\sqrt{\sum_{i=0}^{5} \binom{5}{i}(-\rho^{2} \Delta_\perp)^i}}\Delta_\perp - \frac{\rho^4}{4}\frac{1}{\sqrt{\sum_{i=0}^{5} \binom{5}{i}(-\rho^{2} \Delta_\perp)^i}} \Delta_\perp^2\)\\
  \(4\) & \(\frac{1}{1-\rho^2/2 \Delta_\perp}\)  & \(\frac{1}{1-\Delta_\perp\rho^2/2 }\)  & \(\frac{\rho^2}{2}\frac{1}{\sum_{i=0}^{2} \binom{2}{i}(-\rho^{2}/2\Delta_\perp)^i} \Delta_\perp\) & \(\rho^2\frac{1}{ \sum_{i=0}^{3} \binom{3}{i}(-\rho^{2}/2\Delta_\perp)^i} \Delta_\perp \) 
\end{tabular}
 \end{ruledtabular}
\end{table*}
The Pad{\'e} approximated FLR operators of Table~\ref{table:PadeappstoFLRops} are also utilized for the polarization closures, which are discussed in the next Sec.~\ref{sec:pol}.
\\
Finally, we depict the near Maxwellian FLR operators and their proposed approximations in Fig.~\ref{fig:gamma1234approx}. Here, the Pad{\'e}-approximations qualitatively agree with the square root approximated FLR operators. 
%%%%%%%%%%%%%%%%%%%%%%%%Fig 1
\subsection{Polarization closures}\label{sec:pol}
 %%%%%%%%%%%%%%%exact expression%%%%%%%%%%%%%%%%%%%%%%%%
Analogous to the gyro-average closures of Sec.~\ref{sec:FLR} the polarization closures rely on the near Maxwellian assumption of Eq.~\eqref{eq:F13}. The elementary quantity for the polarization closures is the polarization part of the basic gyro-fluid potential \(\psi_2\) (Eq.~\eqref{eq:psi12def}.
The derivation of the near Maxwellian polarization closure follows the same methodology than that of the near Maxwellian gyro-average closure, but is more involved. 
We calculate \(\psi_2\) by (i) transforming only the gyro-average terms in Eq.~\eqref{eq:Psi12} to Fourier space, (ii) evaluating the \gf moment with the help of the integral identities 
\(\gymom{J_0 ( \varrho K_\perp) }_{13}=e^{-\rho^2 K_\perp^2/2} = e^{-\rho^2(k_\perp^2+k_\perp'^2)/2} e^{-\rho^2 \vec{k}_\perp\cdot \vec{k}_\perp'} \) and \(\gymom{J_0 ( \varrho k_\perp) J_0 ( \varrho k'_\perp) }_{13}= e^{-\rho^2 (k_\perp^2+k_\perp'^2)/2}I_0 (\rho^2 k_\perp k_\perp') \)~\cite{brizard92} where  \(\vec{K}_\perp = \vec{k}_\perp+\vec{k}_\perp'\), 
(iii) Taylor expanding the function
 \(e^{-\rho^2 \vec{k}_\perp \cdot \vec{k}_\perp'} -I_0 (\rho^2 k_\perp k_\perp')=2 \sum_{i=1}^{\infty} \rho^{2i}\left(a_i \mathrm{i}^{2i}  \langle\vec{k}_\perp^{i} , \vec{k}_\perp'^{i}\rangle- b_i  k_\perp^i k_\perp'^i \right) \), 
where 
 \(\langle \mathbb A , \mathbb B \rangle := \mathbb{A}_{i_1,\dots,i_n} \mathbb{B}^{i_1,\dots,i_n}\) fulfills the identity \(\langle \mathbb A , \mathbb A \rangle  = \tnorm{\mathbb A}^2 \) and finally (iv) performing the inverse Fourier transformation. 
 This yields the polarization part of the basic near Maxwellian \gf potential \(\psi_2\), which includes only the square of linear differential operators:
\begin{align}\label{eq:psi2momarb}
  \psi_2 =&  
    \sum_{i=1}^{\infty}  \frac{q  \rho^{2i-2}}{m \Omega^2 } \left[  
    b_{i}\left(\Gb{1}{}{} \Deltabar_\perp^{i/2} \phi\right)^2 -a_{i} \tnorm{\Gb{1}{}{} \nablabar_\perp^i \phi }^2   \right] .
\end{align}
Here, we introduced the Taylor series coefficients
\begin{align}
 a_i &:=  \frac{1}{2 i!},&
 b_i &:= 
  \Bigg\{ \begin{array}{l@{\quad}l} 
 \frac{1}{2^{1+i} (i/2)!^2} & \text{if \(i\) even}  \\  
 0 &   \text{if \(i\) odd}
\end{array}, \nonumber
\end{align}
of the function \(\left[\exp(x^2)-1\right]/2\) and  \(\left[I_0(x^2)-1\right]/2\) around \(x=0\), respectively. The first few coefficients are explicitly given by 
\(a_i=\left\{1/2,1/4,1/12,1/48,\dots \right\}\) and 
\(b_i=\left\{0,1/8,0,1/128,\dots \right\}\). Analogously, to the near Maxwellian gyro-average closure of Eq.~\eqref{eq:gamma1} the near Maxwellian polarization closure of Eq.~\eqref{eq:psi2momarb} reproduces the \fullF  expression of Eq.~\eqref{eq:psi2arb} to \(\mathcal{O}(b^4)\). Consequently, it is hallmarked by \fullF  polarization effects up to  \(\mathcal{O}(b^4)\) 
and near Maxwellian polarization effects for all higher even orders of \(b\).
\\
From the variational of Eq.~\eqref{eq:Npoldef2} with the near Maxwellian polarization closure of Eq.~\eqref{eq:psi2momarb} we 
obtain the near Maxwellian polarization density
\begin{align}\label{eq:Npolarb}
\vec{P}_2 =& 
\sum_\alpha\frac{2 q^2}{m} \sum_{i=1}^{\infty} \Bigg[ \mathrm{i}^{2i} a_{i} (\vec{\nabla} \cdot)^{i-1}  \left(\Gdag{1}{}{}  \frac{\Ngy \rho^{2i}}{\rho^2\Omega^2 }\G{1}{}{} \vec{\nabla}_\perp^{i} \phi \right)
\nonumber \\ &
- b_i \vec{\nabla}_\perp \Delta_\perp^{i/2-1}  \left(\Gdag{1}{}{} \frac{ \Ngy \rho^{2i}}{\rho^2\Omega^2 }\G{1}{}{} \Delta_\perp^{i/2} \phi  \right)\Bigg],
\end{align}
which agrees with the general Laguerre-Hermite expanded expression of Ref.~\cite{frei19} in the near Maxwellian and \(\Bsp\approx B\) limit. 
Here, we introduced the notation \((\vec{\nabla}\cdot)^{i}\) for \(i\)-times a dot product. For example, for \(i=2\) we obtain 
 \(\left(\vec{\nabla} \cdot\right)^2 = \vec{\nabla} \cdot\vec{\nabla} \cdot\).
\subsubsection{Arbitrary order approximation}\label{eq:keqarb}
The near Maxwellian polarization closure of Eq.~\eqref{eq:psi2momarb} together with the higher polarization closures, based on Eqs.~\eqref{eq:Dorlandidentities}, and the polarization density of Eq.~\eqref{eq:Npolarb} are unpractical for numerical computations due to an infinite set of differential operators. In Fourier space this statement converts to an infinite number of convolutions, which pertains also  for the Hermite-Laguerre expanded formulation~\cite{frei19}.
%%%%%%%%%%%%%%%truncated expression%%%%%%%%%%%%%%%%%%%%%%%%
Thus, the remaining task is to find truncations of the infinite series expression of Eq.~\eqref{eq:psi2momarb}, which 
(i) retain the basic quadratic structure of Eq.~\eqref{eq:psi2momarb} for energetic consistency and 
(ii) feature a polarization density with the proper Oberbeck-Boussinesq and \LW limit.
These requirements are fulfilled 
if we truncate at \(K \in \left\{2\mathbb{N}+1 \right\}\cup \left\{2\right\}\) and utilize the Pad{\'e}-approximations
\begin{align}
 \Gsq{1}{1}{2r} := 
  \left\{ \begin{array}{l@{\quad}l} 
 \G{1}{1}{r}& r=2  \\  
\sqrt{\prescript{\scriptscriptstyle 1}{\scriptscriptstyle 2r}{\left(\G{1}{}{}^2\right)} } &   r = \text{odd}
\end{array}\right. 
 \end{align}
 instead of the basic FLR operator \(\G{1}{}{}\). This yields the \(\mathcal{O}\left(b^{2K}\right)\) accurate rational approximation to the polarization part of the basic gyro-fluid potential
\begin{align}\label{eq:psi2momarbtruncN}
 \psi_2 \approx& 
\sum_{i=1}^{K}
\frac{q  \rho^{2i-2}}{m \Omega^2 } \bigg[ b_{i}\left(\Gsqb{1}{1}{2K}\Deltabar_\perp^{i/2} \phi\right)^2 -a_{i} \tnorm{\Gsqb{1}{1}{2K}\nablabar_\perp^i \phi}^2 
   \bigg],
\end{align}
which mimics \AW polarization effects through the \(\G{1}{}{} \approx \Gsq{1}{1}{2K}\) approximation.
Analogously to  Eq.~\eqref{eq:Npolarb} the corresponding \(\mathcal{O}\left(b^{2K}\right)\) approximated polarization density to Eq.~\eqref{eq:psi2momarbtruncN} is derived to
\begin{align}\label{eq:NpolarbtruncN}
  \vec{P}_2 \approx& 
  \sum_\alpha\frac{2 q^2}{m} \sum_{i=1}^{K} \bigg[ \mathrm{i}^{2i} a_{i} (\vec{\nabla} \cdot)^{i-1}  \left(\Gsqdag{1}{1}{2K}  \frac{\Ngy \rho^{2i}}{\rho^2\Omega^2 }\Gsq{1}{1}{2K} \vec{\nabla}_\perp^{i} \phi \right)
\nonumber \\ &
- b_i \vec{\nabla}_\perp \Delta_\perp^{i/2-1}  \left(\Gsqdag{1}{1}{2K}\frac{ \Ngy \rho^{2i}}{\rho^2\Omega^2 }\Gsq{1}{1}{2K} \Delta_\perp^{i/2} \phi \right) \bigg].
\end{align}
\\
Note that the \(\mathcal{O}\left(b^{2K}\right)\) polarization closures retain only \(\mathcal{O}\left(b^{4}\right)\) full-F polarization effects due to the near Maxwellian assumption. 
Thus, we derive in the following explicit and numerically feasible expressions for the truncated polarization closures at \(K\in \left\{1,2\right\}\).
\subsubsection{Second order approximation}\label{eq:keq2}
In the \(K=1\) case Eq.~\eqref{eq:psi2momarbtruncN} reduces to an \(\mathcal{O}(b^2)\) accurate approximation 
\begin{align}\label{eq:psi2momarbtruncG012}
 \psi_2    &\approx  -      \frac{q }{2 m \Omega^2} \tnorm{\Gsqb{1}{1}{2} \nablabar_\perp \phi}^2,
\end{align}
which features arbitrary wavelength polarization effects through the \(\G{1}{}{} \approx\Gsq{1}{1}{2}\) approximation.
%higher closures
The remaining polarization parts of the higher \gf potentials  \(\chi_2,\iota_2\) are consistently derived from the recursive closure formulas of Eqs.~\eqref{eq:Dorlandidentities} together with the polarization part of the basic \gf potential \(\psi_2\) of Eq.~\eqref{eq:psi2momarbtruncG012}:
 \begin{subequations}
\label{eq:chi2G012iota2G012}
\begin{align}
\label{eq:chi2G012}
 %%%%%%%%%%%%%%%%%%%%%%
\chi_2 \approx& - \frac{q }{ m \Omega^2} \Gsqb{1}{1}{2} \nablabar_\perp \phi\cdot \Gsqb{2}{1}{2}  \nablabar_\perp \phi,  \\
  %%%%%%%%%%%%%%%%%%%%%%%%%
  \label{eq:iota2G012}
\iota_2 \approx&  - \frac{q}{m \Omega^2} \left( \tnorm{\Gsqb{2}{1}{2} \nablabar_\perp \phi}^2+
                    \Gsqb{1}{1}{2} \nablabar_\perp \phi
                                        \cdot 
                      \Gsqb{3}{1}{2} \nablabar_\perp \phi
                      \right).
\end{align}
\end{subequations}
The associated \(\mathcal{O}(b^2)\) accurate polarization density
\begin{align}\label{eq:NpolarbtruncG012}
\vec{P}_2 &\approx -\sum_\alpha \frac{q^2}{m}  {\Gsqdag{1}{1}{2}}\frac{\Ngy}{\Omega^2 } \Gsq{1}{1}{2}\vec{\nabla}_\perp \phi 
\end{align}
follows from Eq.~\eqref{eq:NpolarbtruncN}.
The Pad{\'e}-approximated FLR operators appearing in Eqs.~\eqref{eq:psi2momarbtruncG012}-\eqref{eq:NpolarbtruncG012} are to be found in the \(q=2\) row of Table~\ref{table:PadeappstoFLRops}.
\\
We stress that the herein proposed second order polarization charge density  \(-\vec{\nabla} \cdot \vec{P}_2\approx \sum_\alpha \frac{q^2}{m} \vec{\nabla}\cdot \left( {\Gsqdag{1}{1}{2}}\frac{\Ngy}{\Omega^2 } \Gsq{1}{1}{2}\vec{\nabla}_\perp \phi  \right) \) does not agree with the widely used ad-hoc second order Pad{\'e}-approximation in gyro-kinetic models 
\(-\vec{\nabla} \cdot \vec{P}_2\approx \sum_\alpha \frac{q^2}{m} \left[1-\vec{\nabla}\cdot (\rho^2\vec{\nabla}_\perp)\right]^{-1}\vec{\nabla}\cdot\left(\frac{\Ngy}{\Omega^2 } \vec{\nabla}_\perp \phi  \right) \)~\cite{idomura03,bottino04,ku09,dominski17,mishchenko19}, which is not producing a quadratic kinetic \ExB energy. 
Non-linear gyro-kinetic simulations with this ad-hoc Padé-approximation and a full-F arbitrary wavelength treatment of the polarization charge density show only slight differences for typical turbulent transport observables, such as the heat diffusivity~\cite{dominski17}.
\subsubsection{Fourth order approximation}\label{eq:keq4}
In the \(K=2\) case the closure derived from Eq.~\eqref{eq:psi2momarbtruncN} reads
\begin{align}\label{eq:psi2momarbtruncG014}
 \psi_2 
 \approx  - \frac{q }{2 m \Omega^2} \bigg\{&\tnorm{\Gsqb{1}{1}{4} \nablabar_\perp \phi }^2 
 + \frac{ \rho^2}{4}\bigg[2 \tnorm{\Gsqb{1}{1}{4} \nablabar_\perp^2 \phi }^2
 \nonumber \\
 &-\left(\Gsqb{1}{1}{4}\Deltabar_\perp \phi\right)^2 \bigg] \bigg\},
\end{align}
has \(\mathcal{O}\left(b^4\right)\) accuracy and keeps arbitrary wavelength effects due to the \(\G{1}{}{}\approx \Gsq{1}{1}{4}\) approximation.
Analogous to the second order approximation we derive from the
fourth order approximation of Eq.~\eqref{eq:psi2momarbtruncG014} the polarization parts of the higher \gf potentials
 \begin{subequations}
\label{eq:chi2iota2G014}
\begin{align}
\label{eq:chi2G014}
 \chi_2 \approx& - \frac{q }{ m \Omega^2} \bigg\{\Gsqb{1}{1}{4} \nablabar_\perp \phi\cdot \Gsqb{2}{1}{4} \nablabar_\perp \phi
  \nonumber \\   &
  + \frac{\rho^2}{8} \Big[2 \tnorm{\Gsqb{1}{1}{4} \nablabar_\perp^2 \phi}^2 + 4\Gsqb{1}{1}{4}\nablabar_\perp^2 \phi: \Gsqb{2}{1}{4}\nablabar_\perp^2 \phi 
  \nonumber \\  &
  -\left[ \Gsqb{1}{1}{4} \Deltabar_\perp \phi\right]^2  - 2 \Gsqb{1}{1}{4}  \Deltabar_\perp \phi\hspace{1mm}\Gsqb{2}{1}{4} \Deltabar_\perp \phi\Big] 
  \bigg\},
  \\
  %%%%%%%%%%%%%%%%%%%%%%%%%
  \label{eq:iota2G014}
\iota_2 \approx&  - \frac{q}{m \Omega^2} \bigg\{ \tnorm{\Gsqb{2}{1}{4} \nablabar_\perp \phi}^2 +\Gsqb{1}{1}{4} \nablabar_\perp \phi\cdot \Gsqb{3}{1}{4} \nablabar_\perp \phi
  \nonumber \\  &
  +\frac{\rho^2 }{4} \bigg[2 \tnorm{\Gsqb{2}{1}{4} \nablabar_\perp^2 \phi}^2 -\left(\Gsqb{2}{1}{4} \Deltabar_\perp \phi\right)^2 
  + 2\Gsqb{1}{1}{4} \nablabar_\perp^2 \phi 
      \nonumber \\ &
  : 
    \left(\Gsqb{1}{1}{4} \nablabar_\perp^2 \phi + 2 \Gsqb{2}{1}{4} \nablabar_\perp^2 \phi 
    + \Gsqb{3}{1}{4} \nablabar_\perp^2 \phi\right)
  -  \Gsqb{1}{1}{4} \Deltabar_\perp \phi
    \nonumber \\ &
    \times 
  \left(\Gsqb{1}{1}{4} \Deltabar_\perp \phi  
  + 2 \Gsqb{2}{1}{4} \Deltabar_\perp \phi  + \Gsqb{3}{1}{4} \Deltabar_\perp \phi\right)\bigg]
  \bigg\},
\end{align}
\end{subequations}
and the polarization density
\begin{align}\label{eq:NpolarbtruncG014}
\vec{P}_2 \approx-\sum_\alpha \frac{q^2}{m}  \bigg[& \Gsqdag{1}{1}{4} \frac{\Ngy}{\Omega^2 }\Gsq{1}{1}{4}\vec{\nabla}_\perp \phi
\nonumber \\
         &
-2\vec{\nabla} \cdot \left(\Gsqdag{1}{1}{4}\frac{\Ppgy}{4 m \Omega^4}\Gsq{1}{1}{4} \vec{\nabla}_\perp^2 \phi \right)
         \nonumber \\
         &
        + \vec{\nabla}_\perp \left(\Gsqdag{1}{1}{4}\frac{\Ppgy}{4 m \Omega^4} \Gsq{1}{1}{4}\Delta_\perp \phi \right)\bigg].
\end{align}
Again we refer the reader for the various Pad{\'e}-approximated FLR operators in Eqs.~\eqref{eq:psi2momarbtruncG014}-\eqref{eq:NpolarbtruncG014} to the \(q=4\) row of Table~\ref{table:PadeappstoFLRops}.
\\
The fourth order approximation (\(K=2\)) is more complex to implement numerically than the second order approximation (\(K=1\)) due to higher order spatial derivatives, but it does not require to compute the square root of an operator (cf. Table~\ref{table:PadeappstoFLRops}).

In summary, the proposed approximated polarization closures
of Eqs.~\eqref{eq:psi2momarbtruncN}-\eqref{eq:NpolarbtruncN} or more explicitly of Eqs.~\eqref{eq:psi2momarbtruncG012}-\eqref{eq:NpolarbtruncG012} and Eqs.~\eqref{eq:psi2momarbtruncG014}-\eqref{eq:NpolarbtruncG014} are accurate up to \(\mathcal{O}(b^{2K})\), \(\mathcal{O}(b^2)\) and \(\mathcal{O}(b^4)\), respectively. Further they imply 
accurate polarization effects at \AWs in the Oberbeck-Boussinesq limit, the correct \LW limit and produce an appropriate energy conservation law, which is demonstrated in the following. 
%%%%%%%%%%%%%%%%%%%%%%
\subsubsection{Oberbeck-Boussinesq and long perpendicular wavelength limit}\label{sec:OB}
\begin{figure}[ht]
\centering
\includegraphics[width= 0.475\textwidth]{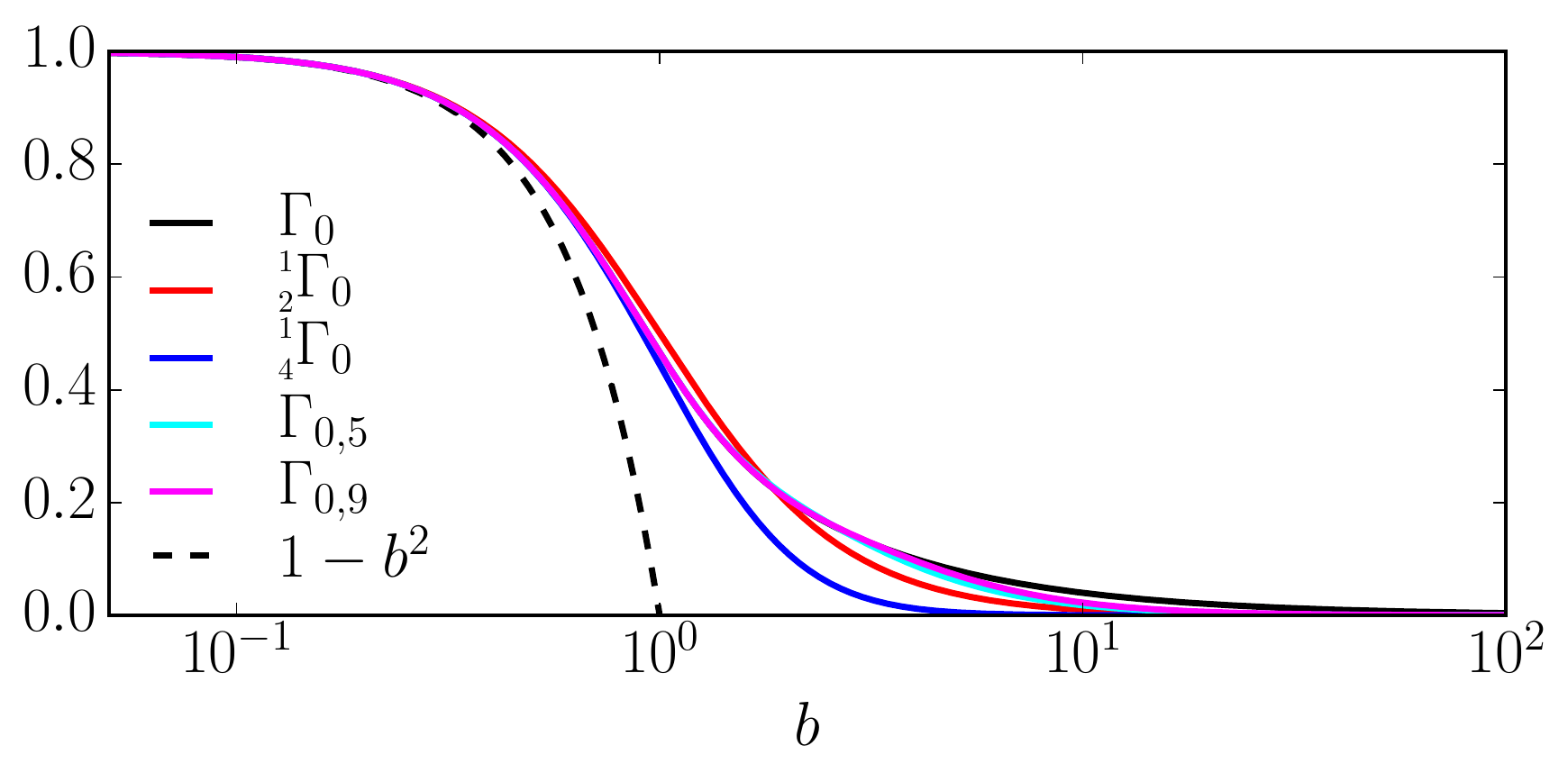}
\caption{Various approximations to the polarization operator \(\G{0}{}{}\) as a function of \(b\) are shown. The \((1,2)\) (red) and \((1,4)\) (blue) Pad{\'e}-approximations agree with  relative errors of 7\% and 11\%, respectively, whereas the \LW approximation fails at \LWsdot 
Higher accuracy is obtained with the higher order rational approximations \(\Gamma_{0,5}\) and \(\Gamma_{0,9}\) (cyan, magenta).}
\label{fig:gamma0approx}
\end{figure}

In the Oberbeck-Boussinesq limit the spatial dependence of \gf moment variables (\(\Ngy,\Ppgy,\dots\)) and the magnetic field magnitude \(B\) is neglected in the polarization closure terms. 
Further, only the stationary contributions of the \gf moment variables  (\(\Ngystat,\Ppgystat,\dots\)), which result from the various \gf moments with the stationary \gy Maxwellian \(\Fstat\), are taken into account.
As a consequence, the polarization and FLR operators are self-adjoint (e.g. \(\G{1}{}{} = \Gdag{1}{}{}\)), commute with spatial derivatives and are to be understood to
contain only contributions from the stationary thermal gyro-radius \(\rhothstat:=\sqrt{\Tpgystat m}/(q B)\).
Thus, the basic near Maxwellian second order polarization charge density resulting from Eq.~\eqref{eq:Npolarb} reduces in the Oberbeck-Boussinesq limit to
\begin{align}\label{eq:NpolarbOBlimit}
- \vec{\nabla} \cdot \vec{P}_2 \approx& \sum_\alpha \frac{q^2 \Ngystat}{m \Omega^2 \rhothstat^2 } \left(\Gstat{0}{}{}-1 \right)  \phi,
\end{align}
where we used \((\vec{\nabla}\cdot)^i\vec{\nabla}_\perp^i = \Delta_\perp^i \).  Note now that the polarization charge density of Eq.~\eqref{eq:NpolarbOBlimit} is similar to the one directly obtained from the \gy Poisson Eq.~\eqref{eq:Poissonvargy} with \(\Fbar \approx \Fstat\).
\\
The Oberbeck-Boussinesq limit of the proposed \(\mathcal{O}\left(b^{2K}\right)\) approximation of Eq.~\eqref{eq:NpolarbtruncN} produces a polarization charge density equal to Eq.~\eqref{eq:NpolarbOBlimit} except that the near Maxwellian polarization operator \(\G{0}{}{}\) is replaced by the truncated polarization operator
\(\Gstat{0,K}{}{} := 1+2(\Gsqstat{1}{1}{2K})^2\sum_{i=1}^{K} \left( b_i-\mathrm{i}^{2i} a_{i}\right)\left(\rhothstat^{2}\Delta_\perp\right)^{i}\nonumber\), 
which converges to \(\G{0}{}{}\) for \(K\rightarrow\infty\).
Strikingly, for the second and fourth order approximations (Eq.~\eqref{eq:NpolarbtruncG012} and Eq.~\eqref{eq:NpolarbtruncG014}) 
the truncated polarization operator resembles the \((1,2)\) or \((1,4)\) Pad{\'e}-approximation of the polarization operator, so that \(\G{0,1}{}{}=\G{0}{1}{2}\) or \(\G{0,2}{}{}=\G{0}{1}{4}\). 
The latter Pad{\'e}-approximations retain high accuracy to the near Maxwellian polarization operator \(\G{0}{}{}\), which is depicted in Fig.~\ref{fig:gamma0approx}. The relative error of the \((1,2)\) and \((1,4)\) Pad{\'e}-approximation to \(\G{0}{}{}\) are of comparable magnitude and are roughly 7\% and 11\%, respectively. Additionally, the \(\mathcal{O}\left(b^{10}\right)\) and \(\mathcal{O}\left(b^{18}\right)\) accurate approximations are shown.

In the \LW limit the presented near Maxwellian and approximated polarization closures reduce to the \LW polarization expressions \(\psi_2 \approx -\frac{q}{2 m \Omega^2} \tnorm{\nablabar_\perp \phi}^2\), \(\vec{P}_2 \approx -\sum_\alpha \frac{q^2 \Ngy }{m \Omega^2 } \vec{\nabla}_\perp \phi\) and \(\chi_2 \approx \iota_2 \approx 0 \)~\cite{madsen13}.
%%%%%%%%%%%%%%%%%%%%%%%%%%%%%%%%%%%%%%%%%%%%%%%%%
\subsection{Gyro-fluid energy conservation}\label{sec:energy}
The proposed \AW polarization closure extension manifests itself only 
in the kinetic \ExB energy part of the total \gf energy conservation law.
The conservation of the total energy \(\frac{d}{d t} E =0\) is obtained 
from the volume integral and species sum over the \gf moment evolution Eq.~\eqref{eq:gfmomevo}, where we replace the phase-space function \(\zeta\) by the Hamiltonian \(H\) of Eq.~\eqref{eq:hamiltonian} and the \gy distribution function \(\Fbar\) by the near Maxwellian \(\Fbar_{13}\) of Eq.~\eqref{eq:F13}. 
This yields the total energy
\begin{align}
\label{eq:energy}
 E &:= \sum_\alpha\int \volXbar\left( \Ppgy  + \frac{\Ppagy}{2}   + \frac{m \Ngy \Ugy^2}{2} -  q \Ngy \psi_2\right), 
\end{align}
which is the superposition of the  internal perpendicular and parallel energy, the parallel kinetic energy and 
the positive definite kinetic \ExB energy
\begin{align}
\label{eq:ExBenergy}
 E_{k} &:=- \sum_\alpha q \int \volXbar \Ngy \psi_2.
\end{align}
In the course of the derivation of energy conservation law we made use of 
 the time evolution equation of the kinetic \ExB energy, which is explicitly computed to
\(  \frac{d E_k}{d t}  = \sum_\alpha q \int \volXbar \left[  \left(\psi -\chi\right)\frac{\partial  \Ngy}{\partial t}  +\frac{\chi}{T_\perp}\frac{\partial \Ppgy}{\partial t} \right] \), 
  the \gf Poisson equation of Eq.~\eqref{eq:gfpoisson} and assumed vanishing surface integral contributions. We stress that two properties must hold for exact conservation of the total energy of Eq.~\eqref{eq:energy}.
First, \(\psi_2\) must be made of squares of linear differential operators, as given by both the near Maxwellian or truncated expression. Second, both \(\chi_1\) and \(\chi_2\) must be derived by the recursive identity of Eq.~\eqref{eq:Dorlandidentities}. 
This enables us to retain energy conservation even if simpler approximations for the remaining higher FLR operators are utilized, for instance the \LW fit \(\Gsq{3}{1}{2}\approx 2\Gsq{2}{1}{2}\) or \(\Gsq{3}{1}{4}\approx 2\Gsq{2}{1}{4}\)~\cite{dorland93}.
Finally, we note that total energy of Eq.~\eqref{eq:energy} agrees in the \LW limit with Ref.~\cite{madsen13}.
%%%%%%%%%%%%%%%%%%%%%%%%%%%%%%%%%%%%%
\section{Conclusions}\label{sec:conclusion}
In this work novel \fullF  \gf closures are derived for polarization effects for arbitrary perpendicular wavelengths, which overcome limitations of former truncated or ad-hoc polarization closures.
Based on a near Maxwellian assumption explicit expressions for the polarization part of the basic \gf potential and the polarization density are derived in Eqs.~\eqref{eq:psi2momarb}-\eqref{eq:Npolarb}. However, these expressions contain an infinite set of spatial operators and require suitable approximations for numerical computations. 
\\
Thus, a general approximation is deduced from a series truncation of the polarization part of the near Maxwellian basic \gf potential of Eq.~\eqref{eq:psi2momarb}, 
which retains the original quadratic structure of the expression for energetic consistency and replaces the inherent FLR operators by appropriate Pad{\'e}-approximations. The resulting Pad{\'e}-based closure for the polarization part of the basic \gf potential of Eq.~\eqref{eq:psi2momarbtruncN} is \(\mathcal{O}\left(b^{2K}\right)\) accurate. Notably, the associated polarization density of Eq.~\eqref{eq:NpolarbtruncN} comprises polarization effects to \AWs since its Oberbeck-Boussinesq limit (Eq.~\eqref{eq:NpolarbOBlimit}) yields an \(\mathcal{O}\left(b^{2K}\right)\) accurate rational approximation to the polarization operator \(\G{0}{}{}\).
%2nd and 4th order
The truncated polarization closures are specified to the tractable limit of second and fourth order accuracy in Eqs.~\eqref{eq:psi2momarbtruncG012}-\eqref{eq:NpolarbtruncG012} respective Eqs.~\eqref{eq:psi2momarbtruncG014}-\eqref{eq:NpolarbtruncG014}, where the polarization parts of the higher \gf potentials of Eqs.~\eqref{eq:chi2G012iota2G012} and Eqs.~\eqref{eq:chi2iota2G014} are consistently closed by the recursive identities of Eq.~\eqref{eq:Dorlandidentities}.
In this limit the rational approximation of the polarization operator \(\G{0}{}{}\) is optimal so that it reduces to its \((1,2)\) and \((1,4)\) Pad{\'e}-approximation.
The truncated near Maxwellian polarization closures ensure energy conservation with a positive definite kinetic \ExB energy as given by Eq.~\eqref{eq:ExBenergy}.
 
%Pade extension
We emphasize that the proposed approximations for gyro-averaging (Table.~\ref{table:PadeappstoFLRops}) and polarization (Eqs.~\eqref{eq:psi2momarbtruncG012}-\eqref{eq:NpolarbtruncG012} or Eqs.~\eqref{eq:psi2momarbtruncG014}-\eqref{eq:NpolarbtruncG014}) are the \fullF  generalization of the widely used \(\dF\) Pad{\'e}-model~\cite{hammett92,dorland93,beer96,snyder01,scott10a}.
%relevance for gyro-kinetic
Finally, the proposed second and fourth order Pad{\'e}-based approximations for \AW polarization effects can be also utilized in gyro-kinetic models, when a (near) Maxwellian \gy distribution function is assumed for the polarization contributions in the \gy Vlasov-Poisson system. 
In particular, we recommend that the ad-hoc second order Pad{\'e}-approximation for \AW polarization effects in gyro-kinetic codes~\cite{idomura03,bottino04,ku09,dominski17,mishchenko19} is replaced with the herein proposed second order Pad{\'e}-approximation of Eqs.~\eqref{eq:psi2momarbtruncG012}-\eqref{eq:NpolarbtruncG012} in order to restore energetic consistency. 
\section{Acknowledgements}
The authors acknowledge helpful discussion with B. J. Frei.
\bibliography{SWLpol_iop_biblio.bib}
\bibliographystyle{aipnum4-1.bst}
\end{document}